\documentstyle[eqsecnum,aps,epsf,preprint]{revtex}
\newcommand{\be}{\begin{equation}}
\newcommand{\bea}{\begin{eqnarray}}
\newcommand{\ee}{\end{equation}}
\newcommand{\eea}{\end{eqnarray}}
\def\[{\left [}
\def\]{\right]}
\def\({\left (}
\def\){\right)}
\def\r{\rho}

\begin{document}

\title{Inhomogeneous Near-extremal Black Branes}

\author{Gary T. Horowitz \thanks{Electronic address:
gary@cosmic.physics.ucsb.edu}
\,and\,\, Kengo Maeda \thanks{Electronic address:
maeda@cosmic.physics.ucsb.edu}}

\address{Department of Physics, University of California,
Santa Barbara, CA. 93106, USA}


\maketitle
\begin{abstract}
{\small It has recently been shown that there exist stable inhomogeneous neutral
black strings in higher dimensional gravity. These solutions were
motivated by the fact that the corresponding homogeneous solutions are unstable.
We show that there exist  new inhomogeneous black string and black
p-brane solutions 
even when the corresponding translationally invariant solutions are 
stable. In particular, we show there exist inhomogeneous near-extremal 
black strings and p-branes. Some of these solutions remain inhomogeneous 
even when
the size of the compact direction (at infinity) is very small.}

\end{abstract}



\section{Introduction}
Higher dimensional generalizations of black holes, such
as black strings and black branes, have played an important role in string
theory. The first 
solutions of this type
were found by assuming translational symmetry along the 
brane~\cite{Horowitz:cd,Duff:1994an}. 
However it was shown by Gregory and Laflamme~\cite{Gregory:vy,Gregory:1994bj}
that many of these
solutions are unstable to linearized perturbations which are 
inhomogeneous along the brane. Although it was originally believed
that the endpoint of this instability (at least in the neutral case) was
individual black holes, it was recently shown that this is incorrect.
Instead, the solution must settle down to stable inhomogeneous black strings
and $p$-branes~\cite{Horowitz:2001cz}.

In this paper we show that there exists a large class of new stable 
inhomogeneous black branes. These solutions are unrelated
to the Gregory-Laflamme instability and exist even when the corresponding
homogeneous solution (with the same mass and charge) is stable. The
new solutions are charged and near their extremal limit. 
Gubser and Mitra~\cite{Gubser:2000ec} conjectured that for a black brane with
a (noncompact)
translational symmetry, the Gregory-Laflamme instability exists precisely
when the brane is thermodynamically unstable.  They
demonstrated numerically that  a certain class of black holes in
anti-de Sitter space were thermodynamically unstable precisely when they were
also unstable against linear perturbations~\cite{Gubser:2000mm}.
Recently, Reall~\cite{Reall:2001ag} provided strong support for this conjecture.
In the cases we consider, the
near extremal homogeneous black branes have positive
specific heat, so they are expected to be stable.
Nevertheless, we find new inhomogeneous
solutions which are also stable.
This shows that there can be more than one stable black
brane configuration 
with the same mass and charge.

We cannot write down the new solutions explicitly. Instead we construct
time symmetric initial data with an apparent horizon whose area is much larger 
than that of the homogeneous black brane with the same mass and charge.
Under evolution, the mass can only decrease (through the emission of radiation)
and the horizon area can  only increase, so the final state will still have
a horizon with area much larger than the corresponding homogeneous solution.
It must be a new stable black brane solution which is not translationally 
invariant along the brane.

To illustrate our construction of the initial data, consider the simple
case of five dimensional Einstein-Maxwell theory. In addition to 
charged black holes with $S^3$ event horizons, this theory is known to have
translationally invariant charged black strings.
To keep the total mass and horizon area finite, let us compactify the
direction along the string so its horizon has topology  $S^2\times S^1$.
Since there is no force between extreme charged black holes, one can
explicitly write down the solution for a one dimensional periodic array,
which is equivalent to a black hole in a space with one direction compactified.
We can now compare the horizon area for the black hole and black string
with the same mass, charge, and length of the circle at infinity.
Interestingly, in the extremal limit,
the black string has zero horizon area while the area of the black hole
remains nonzero. One
can now produce new solutions by starting with the array of
extreme black holes and adding thin neutral black strings to connect
the horizons. From the compactified viewpoint, the black string
wraps around the circle and has its ends stuck on the black hole.
The total increase in mass can be made very small by taking the black string
to be very thin, so the
configuration is near extremality. There is an $S^2\times S^1$ event horizon 
with area comparable to  the extremal black hole. This configuration
is not static and will evolve. But it cannot settle down to the homogeneous
black string since that has very small horizon 
area near extremality. Instead, it
must settle down to a new inhomogeneous near extremal black string. We will 
show that the homogeneous black string has positive specific heat near
extremality, so these
new solutions are not the result of a Gregory-Laflamme instability.

The new solutions have at least two surprising properties. First,
in constructing the periodic array of extreme black holes, the charge $Q$ and
size of the circle at infinity $L$ are independent parameters. In particular,
one can put a large extreme black hole (with horizon area proportional to 
$Q^{3/2}$) inside a space with arbitrarily 
small $L$. The horizon remains a round sphere and is not distorted.
Instead, the size of the circle grows as one comes in from infinity to 
accommodate the black hole. Our construction thus gives an inhomogeneous
near extremal black string in a space with arbitrarily small $L$ at infinity.
It is unlikely that a similar statement is true for neutral black strings.
(However, see \cite{Gubser:2001ac} for evidence that there exist
inhomogeneous neutral black strings even when
the corresponding translationally invariant solutions are stable.)

The second surprising property follows from the fact that extremal black holes
have infinite throats. The thin black string that we add must go all the
way down the throat to reach the horizon. Thus it has infinite length. This
does not contradict our statement that the net increase in mass is very small,
since the mass is infinitely redshifted near the horizon. However the
infinite length implies that the area of the black string horizon is infinite.
We will show that the black string horizon is an apparent horizon for this
initial data, so we have an apparent horizon with infinite area in a 
space with finite total energy! Previously, the fact that extremal
black holes have infinite throats did not seem to have any physical 
consequences since
all matter thrown into the black hole reaches the horizon in finite time.
However, we see that it can have a major effect when one adds the black string.
On the other hand, the
true event horizon will lie outside the apparent horizon and does not
have to go down the throat. Its area will be finite and probably of order 
the area of the black hole.

One can do a similar construction for general black $p$-branes in string theory.
We can start with a periodic array of extremal black branes if their
horizon area is nonzero. Otherwise, we can start
with a periodic array of near extremal solutions. One can then connect them
with thin neutral black strings or even neutral black branes. We will see
that the horizon area can be made larger than the corresponding ``smeared"
homogeneous solution, so it must evolve into a new inhomogeneous near
extremal black brane. 

In recent years the Bekenstein-Hawking entropy of near extremal black holes
and homogeneous black branes has been understood in string theory. (For
a review see \cite{Peet:2000hn}.) It would
be interesting to understand the entropy of these new 
inhomogeneous black branes. We will make some preliminary comments about this
in section V.

This paper is organized as follows: In section II, we review and compare
the black hole and black string solutions of  five dimensional
Einstein-Maxwell theory. In section III, we construct time symmetric
initial data describing a periodic array of extreme black holes connected
by a neutral black string. We also show that this initial data cannot
evolve into the known translationally invariant solution. In section IV,
we extend this construction to higher dimensions and argue that there must
exist inhomogeneous near extremal black p-branes. Finally, section V
contains some further discussion, including comments about the possible
microstates associated with these solutions.

\section{Review of Static Solutions }
The simplest context to describe the new solutions is five dimensional
Einstein-Maxwell theory with action
\begin{eqnarray}
\label{action1}
S=\int \sqrt{-g}\,
\left[\frac{R}{16\pi G}-\frac{1}{4}F_{ab}F^{ab} \right]d^5 x,
\qquad a,b=0,...,4
\end{eqnarray}
where $G$ is the five dimensional Newton constant, $R$ is
scalar curvature, and $F$ is Maxwell field. This theory is known to have
 both electrically charged black holes and translationally invariant,
electrically charged black 
strings. The latter can be viewed as resulting from the collapse of a cylinder
of charged dust. The black holes are described by the five dimensional
generalization of the Reissner-Nordstrom solution. We are mostly
interested in the extremal limit which can easily be obtained as 
follows~\cite{Myers:rx}.
Let
\be\label{extremal}
ds^2 = -U^{-2}(x^i) dt^2 + U(x^i) \delta_{ij} dx^i dx^j
\ee
\be\label{BPS}
E_i =\alpha U^{-1} \partial_i U
\ee
where $\alpha = \pm (3/16\pi G)^{1/2}$.
Then the Einstein-Maxwell equations reduce to just the condition that $U$ be a
harmonic function. If $U$ is the field of a point mass, 
\be
U=1+{\mu\over x_i x^i}
\ee
where $\mu$ is a positive constant, 
then the solution~(\ref{extremal}) describes an extremal black hole. 
The spatial metric has an infinite ``throat" since the proper distance
to $x^i=0$ 
is stretched out infinitely, and near the origin
the area of the three-spheres of constant
radius is almost independent of the radius.
In these coordinates, the event horizon is at $x^i=0$
and has area
\begin{eqnarray}
\label{areabh}
A_{\rm BH}=2\pi^2\mu^{3/2}.
\end{eqnarray}
The ADM mass ${M}$ and charge ${Q}$ are given by
\begin{eqnarray}
\label{masschargebh}
{M}=\frac{3\pi}{4G}\,\mu, \qquad
{Q}=\pm \sqrt{\frac{3\pi}{G}}\pi\mu,
\end{eqnarray}
respectively, where we have simply normalized the charge by $Q= \oint E_i dS^i$.

Solutions describing several extremal black holes are easily constructed
by letting $U$ have several point sources. (These are the analog of the
familiar Majumdar-Papapetrou solutions in $3+1$ dimensions 
\cite{Majumdar,Papapetrou}.)
To compactify one direction, we let $U$ be the field of a one dimensional
periodic array of point masses. The resulting metric can be 
written~\cite{Myers:rx}
\begin{eqnarray}
\label{bhs}
ds^2_5&
=&-U^{-2}(r,z)dt^2+U(r,z)(dr^2+r^2d\Omega^2+dz^2), \nonumber \\
U(r,z)&=&1+\frac{\pi\mu}{Lr}\frac{\sinh 2\pi\frac{r}{L}}
{\cosh 2\pi\frac{r}{L}-\cos2\pi\frac{z}{L}},
\end{eqnarray}
where the coordinate $z$ is periodic with period $L$.
The black hole horizon is located at $r=z=0$, where $U$ diverges.
Expanding $U$ near this point yields
\begin{eqnarray}
\label{expansion}
U(r,z)=1+\frac{\mu}{r^2+z^2}+\frac{\pi^2}{3}\frac{\mu}{L^2}
+O\left(\frac{r^2}{L^2}, \frac{z^2}{L^2} \right).
\end{eqnarray}
So the geometry near the horizon reduces to that of the isolated
black hole. 
For $r \gg L$, we have
\be\label{Uasympt}
U= 1+ {\pi\mu\over r L} + O(e^{-2\pi r/L})
\ee
Note that the inhomogeneity in the $z$ direction falls off exponentially
for large $r$. This is expected since, from a four dimensional viewpoint,
$z$ dependent perturbations act like massive fields.
The ADM mass ${M}$ and charge $Q$ of the compactified
solution are identical with that of the single black hole~(\ref{masschargebh}).

It is interesting to note that $\mu$ and $L$ are independent parameters
in this solution: One can fit an
arbitrarily large charged black hole into a space with one direction
compactified on an arbitrarily small circle (at infinity).
This is possible since the
size of the circle depends on $U$. It follows from (\ref{Uasympt})
that when $r\sim L$ the proper length of the circle is of order $\mu^{1/2}$,
independent of $L$.

To obtain the (translationally invariant) charged black string solution
we can proceed as follows. Consider
the following metric and two form
\begin{eqnarray}
\label{metricf}
ds^2_5=e^{-4\phi/\sqrt{3}}dz^2+
e^{2\phi/\sqrt{3}}g_{\mu\nu}dx^\mu dx^\nu, \qquad \mu,\,\nu=0,...,3
\nonumber
\end{eqnarray}
\begin{eqnarray}
\label{twoform}
F_2=\frac{1}{2}F_{\mu\nu}dx^\mu\wedge dx^\nu,
\end{eqnarray}
where $g_{\mu\nu}$ is a four-dimensional metric and $z$ is again a compactified
extra-dimension with period $L$. As shown in \cite{Gibbons:1994vm},
if all functions
are independent of $z$, the action~(\ref{action1}) is reduced to
the following four-dimensional action
\begin{eqnarray}
\label{action2}
S=L\int \sqrt{-g}\,
\left[\frac{1}{16\pi G}[R-2\,(\nabla\phi)^2]-\frac{1}{4}
e^{-(2\phi/\sqrt{3})}F_{\mu\nu}F^{\mu\nu} \right]d^4 x.
\end{eqnarray}
The electrically charged spherically symmetric black hole
solution of this theory is
given by \cite{Garfinkle:qj,Gibbons:1987ps}
\begin{eqnarray}
\label{4bs}
ds^2&=&-\lambda^2 dt^2+\lambda^{-2}dr^2+f^2d\Omega^2, \nonumber \\
\lambda^2&=&\left(1-\frac{r_+}{r}\right)
\left(1-\frac{r_-}{r} \right)^{1/2}, \nonumber \\
f&=&r\left(1-\frac{r_-}{r} \right)^{1/4},
\end{eqnarray}
where $r_+$ and $r_-~(\le r_+)$ are free parameters.
The Maxwell field $F$ and the ``dilaton field" $\phi$ are
\begin{eqnarray}
\label{Mfield}
F_2=\pm \frac{1}{4r^2}\sqrt{\frac{3r_+r_-}{\pi G}}dt\wedge dr
\end{eqnarray}
and
\begin{eqnarray}
\label{dilaton}
e^\phi=\left(1-\frac{r_-}{r}\right)^{\sqrt{3}/4},
\end{eqnarray}
respectively.
Substituting Eqs.~(\ref{4bs}) and (\ref{dilaton}) into Eq.~(\ref{metricf}),
we obtain the five-dimensional black string solution:
\begin{eqnarray}
\label{bss}
ds^2_5=-\left(1-\frac{r_+}{r}\right)\left(1-\frac{r_-}{r}\right)dt^2
+{\left(1-\frac{r_+}{r}\right)}^{-1}dr^2
+r^2\left(1-\frac{r_-}{r}\right)d\Omega^2+
{\left(1-\frac{r_-}{r}\right)}^{-1}dz^2.
\end{eqnarray}
The event horizon is at $r=r_+$ and has area
\be\label{horarea}
A=4\pi r_+^2 L \(1-{r_-\over r_+}\)^{1/2}.
\ee
There is a curvature singularity at $r=r_-$. In the extremal limit, $r_+
\rightarrow r_-$, the horizon area clearly goes to zero. By a simple
coordinate transformation, the extremal
solution can be put into the form~(\ref{extremal}), where now $U$ is the field
of a line source.
The total mass $\tilde M$ and charge $\tilde Q$ is given by
\begin{eqnarray}
\label{masschargebs}
\tilde M=\frac{L}{4G}(2r_++r_-), \qquad
\tilde Q=\pm L\sqrt{\frac{3\pi r_+r_-}{ G}}.
\end{eqnarray}
In terms of $\tilde Q$ and the Hawking temperature
\begin{eqnarray}
\label{temp}
T_H=\frac{1}{4\pi r_+}\sqrt{\frac{r_+-r_-}{r_+}}
\end{eqnarray}
we can calculate the specific heat $C$ defined as
\begin{eqnarray}
\label{defsh}
C(\tilde Q,T_H)=\frac{\partial M(\tilde Q,T_H)}{\partial T_H}.
\end{eqnarray}
In the near-extreme limit~($r_+\to r_-$),
we obtain the asymptotic form of $C$:
\begin{eqnarray}
\label{sh}
C(\tilde Q,T_H)=\sqrt{\frac{16\pi G}{27}}
\frac{ |\tilde Q|^3}{L^2}\Bigl(T_H+O(T_H^3) \Bigr).
\end{eqnarray}
Since the specific heat $C$ is clearly positive in the near-extreme case,
the solution~(\ref{bss}) is believed to be stable.

Equating the mass of the black hole to that of the black string in
the extremal limit $r_+=r_-$, we find $\mu=Lr_+/\pi$. It then follows
from Eqs.~(\ref{masschargebh}) and (\ref{masschargebs}) that the charges of
the two systems are also equal. But we have seen that the horizon areas 
are very different. We will use this fact in the next section 
to construct initial data for a new inhomogeneous near extremal black string.

\section{Initial data for an inhomogeneous black string}
In this section we show that one can start with the extremal black hole solution
in a space with one direction periodically identified, and add a small amount
of energy so that the horizon changes from having topology $S^3$ to having
topology $S^2 \times S^1$. The idea is to add a thin neutral black string which
goes around the circle and has its ends stuck on the black hole. This configuration
will not be static and will evolve. We do not know the final
exact static solution, but we will construct time symmetric initial data and
then argue that it must settle down to a new inhomogeneous near extremal black string.
There are in fact two different ways to construct the initial data. 
The first is more
explicit, and the second is more general but (at the moment) less rigorous.
We will concentrate on the first method and comment on the second at the
end of this section.

\subsection{Conformal rescaling construction}

Let us consider the following four-dimensional spatial metric:
\begin{eqnarray}
\label{initialdata}
ds^2_4&=&u\left[\Bigl(1-\frac{r_0}{r}\Bigr)^{-1}dr^2
+r^2d\Omega^2+dz^2\right] \nonumber \\
&=&u\,\overline{ds^2},
\end{eqnarray}
where $u$ is a function of $(r,z,\theta,\phi)$.  $z$ is a periodic coordinate
with period $L$. In the case of $u=1$,
the metric represents the spatial part of the neutral black string solution
with Schwarzschild radius  $r=r_0$, while the $r_0=0$ case reproduces
a spatial slice in the extremal solution~(\ref{extremal}).

Assuming time
symmetric initial data, the constraint equations reduce to
\begin{eqnarray}
\label{constraint1}
R=8\pi G E_i E^i
\end{eqnarray}
\begin{eqnarray}
\label{constraint2}
\nabla_i E^i=0, 
\end{eqnarray}
$\nabla$ is the covariant derivative
with respect to the metric $ds_4^2$.
Since the scalar curvature $\bar{R}$ of the metric
$\overline{ds^2}$ is equal to zero,
$R$ can be simply written in terms of $u$:
\begin{eqnarray}
\label{R-u}
R=\frac{3}{2}\frac{(\bar{\nabla}u)^2}{u^3}
-3\frac{\bar{\nabla}^2 u}{u^2},
\end{eqnarray}
where $\bar{\nabla}$ represents the covariant derivative operator of
the metric $\overline{ds^2}$. Setting
\begin{eqnarray}
\label{electric}
E_i=\alpha u^{-1}\bar\nabla_i u,
\end{eqnarray}
where $\alpha$ is a constant, the constraint
Eq.~(\ref{constraint2}) is reduced to
the Laplace equation
\begin{eqnarray}
\label{laplace}
\bar\nabla^2 u=0.
\end{eqnarray}
Substituting Eq.~(\ref{electric})
into Eq.~(\ref{constraint1}), we can obtain $\alpha$ as
\begin{eqnarray}
\label{alpha}
\alpha=\pm\sqrt{\frac{3}{16\pi G}},
\end{eqnarray}
So Eq.~(\ref{laplace}) is the only constraint equation for the
time symmetric initial data.

Let $u$ be the solution with a point source at $\theta=0, r=r_0, z=0$.
This corresponds to adding an extreme
black hole at the north pole of
the black string. Introducing a new coordinate
$r_*=\int^r_{r_0} dr/\sqrt{1-r_0/r}$, we can make an analytic
continuation to the region $r_*<0$. Hence this initial data
has a symmetry $r_*\leftrightarrow -r_*$.
Since the initial data is time symmetric,
the outgoing null expansion $\theta_+$ for any closed three-dimensional
hypersurface is given by the extrinsic curvature of the hypersurface
on the initial data.
This means that $r_*=0~(r=r_0)$ is an apparent horizon. To estimate
the area, let us consider the local geometry around the north pole located
at $r_*=0$ and $\theta=0$. Since the spatial metric $\overline{ds^2}$
is locally flat, $\overline{ds^2}$ is approximately
\begin{eqnarray}
\label{appro}
\overline{ds^2}\sim d\tilde{x}^2+
d\tilde{y}^2+dz^2
+dr_*^2,
\end{eqnarray}
where $(\tilde{x}, \tilde{y})$ are local orthogonal coordinates
around the north pole. The approximate solution of $u$ becomes
\begin{eqnarray}
\label{approU}
u\sim \frac{{\mu}}{\tilde{R}^2}\qquad \mbox{for }\qquad
\tilde{R}^2 \equiv \tilde{x}^2+\tilde{y}^2+z^2+r_*^2\ll r_0,
\end{eqnarray}
where ${\mu}$ is a positive constant proportional to the charge.
The area of the apparent horizon around the north pole is
approximately given by
\begin{eqnarray}
\label{area r^*}
A\sim \int \frac{d\tilde{R}}{\tilde{R}}\Biggr|_{r_*=0}\rightarrow \infty.
\end{eqnarray}
This shows that the apparent horizon~($r=r_0$) has an infinite
area! Since the apparent horizon has $S^1\times S^2$ topology,
the event horizon also has $S^1\times S^2$ topology~\footnote
{Strictly speaking, the topology of the apparent horizon is
$S^1\times S^2$ minus a point $p$. 
However, one can modify the horizon so that the topology becomes
$S^1\times S^2$ by covering the neighborhood of $p$ with a very shallow
cap. As easily checked, the outgoing null expansion
on the cap cannot be positive since $u$ rapidly decreases toward the
outside. This produces a compact apparent horizon with arbitrarily large
(but finite) area.}.

At first glance,
the area of the event horizon
also seems to be infinite because the event horizon is
outside the apparent horizon. This speculation is obviously wrong because
outside of the apparent horizon
$u$ is finite everywhere on the hypersurface.
Although the exact location of the event horizon is unknown, we can
still produce a lower bound on its area. The idea is to
consider a region of the space and show that all surfaces passing through
this region have area larger than some constant. Our estimate will be
crude, but sufficient to establish the existence of new inhomogeneous
solutions.

\begin{figure}
 \centerline{\epsfxsize=10.0cm \epsfbox{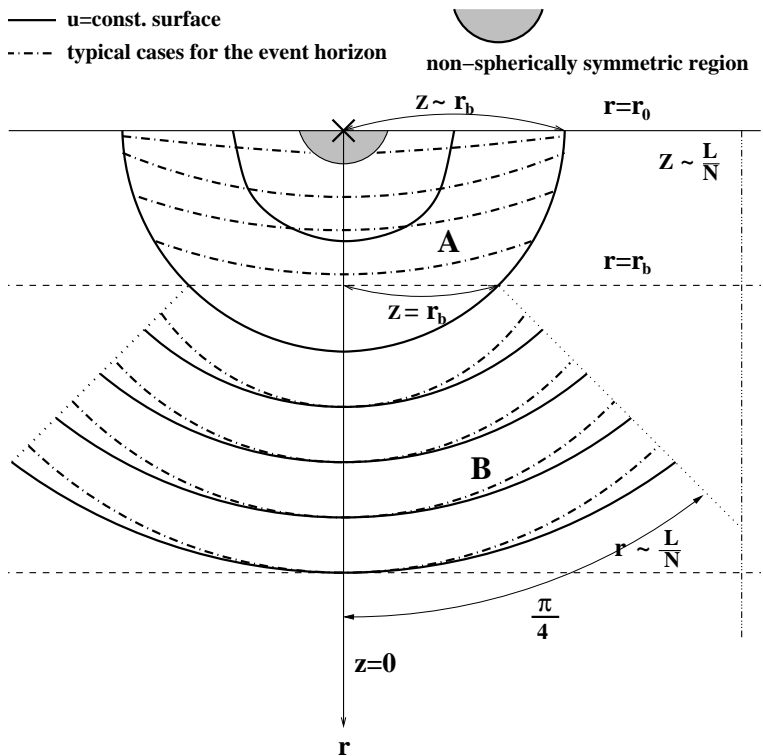}}
      \caption{Constant $u$ surfaces and possible cases for the event
horizon are depicted on the $(r,z)$ surface. Each point represents
a two-sphere. For the region B, the event horizon would be more
convex than the constant $u$ surfaces because the outgoing null
expansion on those surfaces is almost zero due to the infinite throat.
(The outgoing null expansion along
the event horizon should be strictly positive when the spacetime is not
static).}     \protect
\label{nearex}
\end{figure}

Adding the extreme black hole at the north pole of the black string
horizon produces initial data
which is approximately spherically symmetric for $r\gg r_0$. It will
be convenient to have initial data which is approximately spherically
symmetric even for $r \sim 2r_0$. This can be obtained by distributing
a large number $N$ of 
extreme black holes each with charge
$2\,{Q}/{N}>0$ uniformly on the sphere $(r=r_0, z=0)$~
\footnote{Because our initial data has two asymptotically flat regions,
half of the total electric flux reaches each one. Therefore,
we put $2\,{Q}$ on the sphere $(r=r_0,z=0)$ as a total charge so that the total flux 
for one asymptotically flat region is just ${Q}$. }.
We now construct a region such that the effects of both the compactified
extra dimension and the tiny black string are negligible. To this end,
let $r_b = N r_0$. By choosing $r_0$ small enough we can arrange for 
$r_b < L/N$.
Introducing a new radial coordinate $R^2 = r^2 + z^2$ and angular
coordinate $\chi$
such that $r=R\cos\chi,\,z=R\sin\chi$, define region B by
$(R,\xi)\in [\sqrt{2}\,r_b\le R\le L/N,\,
-\pi/4\le \chi\le \pi/4]$~(see Fig.~\ref{nearex}).
Then, the approximate solution in Eq.~(\ref{expansion})
is valid everywhere in $B$.
Also define
region A by the almost all spherically symmetric region enclosed by
the $u={\rm const.}$ surface intersecting a point
$(r=r_b,\, z=r_b)$~(see Fig.~\ref{nearex}).
Since the electric field normal to each $u={\rm const.}$ surface should be positive,
i.e. $E_i\, n^i\sim -(\bar\nabla_i u)\, n^i>0$,
(where $n^i$ is a unit outward normal vector for each $u={\rm const}.$ surface),
$u$ must be increasing as one moves in.
So the following lower bound on $u$ is obtained:
\begin{eqnarray}
\label{lower bound u}%
u|_{A}\ge u|_{(r,z)=(r_b,r_b)}=u_b\sim \frac{\mu}{r_b^2}
\end{eqnarray}
by Eq.~(\ref{expansion})~(see also Fig.~\ref{nearex}).

Now, one can estimate a lower bound $A_0$ for the area of the event horizon.
Fig.~\ref{nearex} shows typical cases for the event horizon passing through
our regions $A$ or $B$.
In general it is possible that the horizon is more complicated and wavy.
In that case the area would be greater than the lower bound of the 
typical cases.
So, it is enough for the estimation of $A_0$ to consider only the typical cases.
Recalling the spatial metric Eq.~(\ref{initialdata}), let us first
consider the case that the event horizon intersects the region A.
Then, the area of the event horizon in this region $A_0$ is roughly
\begin{eqnarray}
\label{lower bound S_A}
A_0\sim
\left(\int_{A} (\sqrt{u}r)^2\sqrt{u}\,dz \right)
\ge \left(\int_{A} (\sqrt{u_b}r)^2\sqrt{u_b}\,dz \right)
\ge \mu^{3/2}
\left(\frac{r_0}{r_b}\right)^2=\frac{\mu^{3/2}}{N^2},
\end{eqnarray}
where we used $r_0/r_b=1/N$.
Note that this lower bound is independent of
$r_0$. In the case that the event horizon intersects the region B,
we can easily obtain a lower bound on $A_0$.  Eq.~(\ref{expansion})
is the approximate solution for the region B, and we will assume the constant
term is negligible. (If necessary, we can choose $N$ so that $\mu/L^2 \gg 1/N^2$
to insure this.) Then the
area of each $u={\rm const}$. surface is almost
$\mu^{3/2}$. So, if the orbit of the event horizon is $R=R(\chi)$,
\begin{eqnarray}
\label{lower bound S_B}
A_0 &\simeq& 4\pi
\int_{-\frac{\pi}{4}}^{\frac{\pi}{4}}
 (u R^2)^{\frac{3}{2}}
\left(1+\left(\frac{R'}{R}\right)^2 \right)^{\frac{1}{2}}
\cos^2\chi\, d\chi \nonumber \\
&\ge& 4\pi
\int_{-\frac{\pi}{4}}^{\frac{\pi}{4}}
 (u R^2)^{\frac{3}{2}}\cos^2\chi\, d\chi
\sim \mu^{\frac{3}{2}},
\end{eqnarray}
where a dash denotes the derivative with respect to $\chi$.
We used Eq.~(\ref{expansion}) to derive the final value.
Finally, suppose the event horizon passes outside of the region B,
and consider the subset with $-L/N < z < L/N$.
In this region $u$ is given by (\ref{bhs}). From this explicit solution
one can show that
$u r^2> \mu$. So the proper area of the two-spheres in this region
is at least $\mu$. Since the length in the $z$ direction is at least $L/N$
(since $u>1$ everywhere) we obtain the lower bound $A_0 > \mu L/N$.
Thus, we can get the lower bound for the area of the event horizon
as
\begin{eqnarray}
\label{boundA}
A_0> \mbox{min}\left\{\frac{\mu^{3/2}}{N^2},\,\mu \frac{L}{N}\right\},
\end{eqnarray}
which is independent of $r_0$.

\subsection{Evolution of initial data}

The initial data constructed above
must evolve since, e.g., the surface gravity is not
constant over the horizon. It is large near the thin black string and
small near the extremal black hole. More physically, one would expect
the extremal black hole to start swallowing up the string. What will it
evolve to?
We will assume cosmic
censorship holds, since a generic violation of cosmic censorship in
five dimensions would be an even more surprising conclusion
than we find below. We will also use the result of \cite{Horowitz:2001cz} that
cycles on an event horizon cannot shrink to zero size. This means
that if the event horizon is initially $S^2\times S^1$, it must
remain $S^2\times S^1$. It cannot pinch off.

The ADM mass of our initial data set is easily computed from 
(\ref{initialdata}). Since $u$ is a solution to Laplace's equation
in the black string background 
with a point source, the asymptotic form of $u$ can be directly related
to the source by integrating the equation over all space. This
means that the asymptotic form is independent of $r_0$ and the same as
(\ref{Uasympt}):
$u\simeq 1+\pi\mu/Lr$. 
The ADM mass is thus
\begin{eqnarray}
\label{ADM massf}
M_{ADM}={M}+\frac{r_0 L}{2G}.
\end{eqnarray}
where $M$ is related to $\mu$ by (\ref{masschargebh}). The total charge $Q$ is
also related to $\mu$ by (\ref{masschargebh}).
So the mass is above the extremal limit just by the second term which
can be made very small by choosing $r_0$ small. Under evolution,
the mass can only decrease  since energy can be radiated away to 
infinity. Since the charge is conserved,
the final state must be even closer to extremality.
On the other hand, the
horizon area can only increase during the evolution
by the usual area theorem. Thus the
final state must be a black string with horizon area larger than the
initial area (\ref{boundA}). But a homogeneous black string  (\ref{bss}) with
small excess energy $r_0 L/2G$ above extremality would have $r_+ - r_- = 4r_0$.
Thus its horizon area (\ref{horarea}) would be $A_0\propto r_0^{1/2}$.
By choosing $r_0$ sufficiently small, this can clearly
be made much smaller than 
the initial horizon area. So the initial data
cannot settle down to the
homogeneous black string. It must approach a new stable, inhomogeneous 
near extremal black string.

We have been discussing classical solutions, but if one has to take $r_0$
less than the Planck scale (or string scale) the solutions would not
be physically interesting. Quantum effects would certainly be important near
the horizon. Fortunately, it is easy to see that this is not necessary.
For our lower bound on the horizon area, $N \sim 100$ should be more than
adequate. So as long as $\mu$ and $L$ are much larger than the Planck scale,
we can choose $r_0$ 
greater than the Planck 
scale and still deduce the existence of new solutions.

It is plausible that the final configuration will resemble a  large 
near extremal
black hole with a near extremal black string going around the 
circle\footnote{We thank S. Ross for suggesting this.}. This
configuration would have low surface gravity everywhere. If
one starts with this inhomogeneous black string and takes its extremal limit,
it is plausible that the solution degenerates to an extremal black hole
with a (singular) extremal black string going around the circle. This
solution can be constructed explicitly. Recall the general form of static
extremal solutions (\ref{extremal}). 
The solution we want is obtained by letting $U$
be the harmonic function with a line source and a point source superposed
on the line. The line source produces the extremal black string, while the
point source reproduces the extremal black hole.
In fact, there is a large class of extremal solutions of this
type, since one can specify the line density arbitrarily.
It is not clear whether there are near extremal analogs of all of these
extremal solutions, or just a few. If all of these extremal solutions can
be obtained as the extremal limit of some black string, then the structure
of black string solutions would be very rich indeed. It would depend on 
an arbitrary function.

\subsection{Gluing construction}

The previous construction of time symmetric initial data is very explicit,
but restricted to situations where one wants to add black strings to extremal
black holes. In the next section we will want to add black strings to near extremal
black p-branes. To do this we need a more general construction which we describe
here. This construction is less explicit (and, at the moment, less rigorous)
but can be used to insert a thin black 
string into essentially any initial data set.

Intuitively, it seems clear that one should be able to take any smooth initial
data set and insert a very small Schwarzschild black hole. This is because every
curved space is locally flat, and the Schwarzschild black hole is approximately flat
after a few Schwarzschild radii from the black hole. (Near the black hole there is
large curvature, but the constraint equations are satisfied.) This has indeed 
been proven
in $3+1$ dimensions in a recent paper \cite{Isenberg:2001qe}.
We would like an analogous result
for black strings in $4+1$ dimensions. In other words, it should be possible to 
remove a neighborhood of a geodesic in an initial data set and glue in 
a thin black string. This possibility is supported by the fact that
the argument in \cite{Isenberg:2001qe} can be easily generalized to show that
one can glue small black holes in $4+1$ dimensions \cite{Isenberg}. But
if one adds a sequence of closely spaced black holes, there will be a single
apparent horizon enclosing all of them which is essentially a black string.

In the case of interest here, we start
with the exact solution describing a large five dimensional
extremal black 
hole with the $z$ direction compactified. The curvature on a static
slice is everywhere small. The $z$ axis comes up the infinite throat of
the extreme black hole,
goes around the circle and then back
down the infinite throat. We now want to remove a neighborhood
of this curve and glue in a thin black string, i.e., $4D$ Schwarzschild
cross ${\bf R}$. The fact that the initial data contains an electric field should
not be a problem. Let the electric field along the $z$ axis be
$E_z(z)$. Since the electric
field is parallel to the $z$ axis, when we glue in the black string, we can
add an electric field (pointing in the ${\bf R}$ direction) equal to $E_z(z)$.

Although we cannot write down this initial data explicitly,
we can write down a metric which describes the solution asymptotically down
the throat of the extreme black hole. Before we add the black string,
the asymptotic solution is just $S^3 \times {\bf R}$
with a constant electric field 
directed up the throat. So
the constraint equation is just that the scalar curvature is a positive
constant. Adding
the black string on either side of the throat, corresponds to adding
two small black holes at the poles of the large $S^3$.
But this is exactly the geometry of a static slice of Schwarzschild de Sitter 
in $3+1$
dimensions. This metric has constant positive scalar curvature by virtue of the
Einstein constraint equation with a positive cosmological constant.
So the initial data down the throat of the extremal black hole
with a black string added is just the product of a 
line and a static slice of $3+1$ Schwarzschild de Sitter. The electric field
is still constant and directed along the line.

The resulting initial data differs from the one obtained by the conformal approach
in the following respect.  In the conformal approach, the size of the black string
apparent horizon
is rescaled by $u$. Thus, near the extremal black hole it becomes much larger than
$r_0$. In fact, it is easy to see that the $r=r_0$ surface intersects the extremal 
black hole horizon at the equator. So the black string has grown to a size given by
$M$! In the gluing approach this does not happen.
The size of the black string stays small,
$O(r_0)$, all the way down the throat.

\section{Generalization to p-branes}

\subsection{Starting with extremal branes}

The above  construction can be  applied to a variety of higher dimensional 
situations to deduce the existence of inhomogeneous near extremal p-branes.
First we place the example we have already discussed in the context of string
theory. Five dimensional Einstein-Maxwell theory
arises from ten dimensional (Type II) string
theory by compactifying on $T^5$ and considering three fundamental charges. 
The standard choice is D1, D5 and momentum\footnote{Starting with this
extremal solution, one can 
add a traveling wave along the D1 direction \cite{Cvetic:1995bj}.
However the local horizon geometry remains homogeneous and is independent
of the wave profile \cite{Horowitz:1996th}.}.
However, due to the momentum,
the ten dimensional metric has an ergoregion which complicates the construction.
This can be avoided by applying a T-duality, so that the solution contains
D0, D4, and fundamental string charge.
If we let the charges all be 
different,
the extremal solution has the following (string frame) metric 
\cite{Tseytlin:1996bh}: 
\begin{eqnarray}\label{threecharge}
ds^2 =& -H_0^{-1/2} H_4^{-1/2}H_1^{-1}dt^2 +
H_0^{1/2} H_4^{-1/2}(dy_i dy^i)\\ \nonumber
   &+ H_0^{1/2} H_4^{1/2} (H_1^{-1} dx^2 +
dr^2 + r^2 d\Omega +dz^2)
\end{eqnarray}
where 
\be
H_0 = 1+ {\mu_0\over r^2 + z^2}, \qquad H_1 = 1+{\mu_1\over r^2 + z^2},
\qquad H_4 = 1+ {\mu_4\over r^2 + z^2}
\ee
are three harmonic functions, and $y_i$ are coordinates  on $T^4$.
The 4-branes are wrapped around this $T^4$, and the fundamental strings
are wrapped in the $x$ direction which we also assume is compact. The branes
are all distributed uniformly over these compact directions.
There are four directions transverse to
all the branes labelled by $z,r,\theta,\phi$. There is an event horizon
at the origin of these coordinates, $z=r=0$, with nonzero area. Since this event
horizon is extended in the $x,y_i$ directions, the solution describes
an extremal black five-brane. Its horizon has topology $S^3\times T^5$.
If we set $\mu_0=\mu_1=\mu_4$, then the solution
(\ref{threecharge}) reduces to the product of (\ref{extremal}) and a flat $T^5$.

Since $H_i$ are just harmonic functions, we can compactify the $z$
direction by taking a periodic array as before. The 
homogeneous solution with the same mass and charges
can be obtained by choosing the harmonic functions
to be independent of $z$,
 $\tilde H_i = 1 + \tilde \mu_i/r$. The near extremal solution
is obtained by adding a factor of $(1-r_0/r)$ to $g_{tt}$ and $(1-r_0/r)^{-1}$
to $g_{rr}$ as usual. It is easy to verify
that the near extremal homogeneous solution has positive specific heat,
so by the Gubser-Mitra conjecture \cite{Gubser:2000ec}
it should be stable. Nevertheless,
we now show that there exist near extremal solutions which are inhomogeneous
in the $z$ direction.

To begin, note that the horizon of the homogeneous solution becomes
singular (zero area) in the extremal limit.
This is expected since if the solution is homogeneous in the $z$ direction
also, it can be dimensionally
reduced to a four dimensional black hole. But it is well known that
extremal black
holes in four dimensions will have nonzero area only if there are four nonzero
fundamental charges \cite{Maldacena:1996gb,Johnson:1996ga}. 
Since we have only three nonzero charges, 
the four dimensional extremal black hole
must have zero area. 

Now, starting with the array of extremal black five branes in ten
dimensions, one has several possibilities to construct initial data
that will evolve into a new inhomogeneous black brane solution.
The basic idea is the same as before. By adding a small amount of energy
we can cause the horizon to extend in the $z$ direction. Since the
initial horizon area is much larger than the corresponding near extremal
homogeneous solution, the final state must be something new.

In ten dimensions, there are neutral black $p$-branes which are just
the product of ${\bf R}^p$ and the $(10-p)$-dimensional Schwarzschild solution.
We can use any of the solutions with $1\le p\le 6$ to connect
the extremal black five-branes in the periodic array. For example,
one possibility is to add a thin neutral black string in ten dimensions
which goes around the $z$ direction and ends on the black five-brane.
This will produce an apparent horizon with topology $S^3\times T^5$ with
a $S^7\times {\bf R}$ handle 
added\footnote{That is, remove two small balls from 
$S^3\times T^5$. The boundary of each ball is an $S^7$. Now identify these two 
boundaries.}. Under evolution, the topology of the apparent horizon can 
change, so the final topology of the event horizon may be different. 
But the solution cannot
settle down to the known homogeneous solution.
Another possibility is to use a thin neutral black six
brane to connect the extremal elements in the array. 
(This is just the lift of the black string we added in five dimensions
in the previous section.) This will produce an apparent horizon with topology
$S^2 \times T^6$, which is the same topology as the horizon in
the near extremal homogeneous solution.
But once again, the initial area is too large for the solution to become
homogeneous. Clearly,  one can do a similar construction with any
of the thin neutral black $p$-branes  with $1\le p\le 6$.

One should be able to  use either the gluing approach or a generalization of the
conformal approach (where one rescales different parts of the spatial metric
by different conformal factors) to construct
suitable initial data.

\subsection{Starting with near-extremal branes}

The above solutions contained three nonzero charges in ten dimensions.
If one starts with just a single charge, then the horizon area always
goes to zero in the extremal limit. For example,
consider a near-extremal D3-brane wrapped on a $T^3$ of volume $V$:
\be\label{threebrane}
ds^2 = H_3^{-1/2}(\r)\[-\(1-{\r_0^4\over \r^4}\)dt^2 + dy_i dy^i\] 
+ H_3^{1/2}(\r)
\[\(1-{\r_0^4\over \r^4}\)^{-1} d\r^2 + \r^2 d\Omega_5\]
\ee
where $H_3(\r) = 1 + \mu_3/\r^4$. The horizon is at $\r=\r_0$ and has
area $A=\pi^3\r_0^5
V H_3(\r_0)^{1/2}$. This area goes to zero in the extremal limit ($\r_0 =0$,
$\mu_3$ fixed) so if we start with a one dimensional array of extremal three-branes
and then add neutral black strings, the area of the apparent horizon will 
still be small and not obviously larger than the corresponding homogeneous
solution. 
However, we can still show the existence of inhomogeneous solutions if
we start with an array of near-extremal 3-branes. Such solutions are not
known explicitly, but if the separation between the branes $L$ is
large compared to $\r_0$, it is plausible that one can approximate
the solution by starting with (\ref{threebrane})  and letting $H$ be
the harmonic function on flat ${\bf R}^6$ with a one dimensional array of
point sources separated by $L$. Near each brane, the metric will reduce to the
single brane solution (\ref{threebrane}), and for $r_0=0$, the metric
reduces to the known array of extremal three branes. Let us label
the direction along the array by $z$.

We now want to compare the horizon area of this near extremal array with
that of a solution with the same mass and charge which is uniformly smeared 
over the $z$ direction. That solution is easily obtained by  starting with
the near extremal $D4$-brane and applying T-duality along one direction
of the brane:
\be\label{fourbrane}
ds^2 = \tilde H_3^{-1/2}(r)\[-\(1-{r_0^3\over r^3}\)dt^2 + dy_i dy^i\] 
+ \tilde H_3^{1/2}(r)
\[\(1-{r_0^3\over r^3}\)^{-1} dr^2 + r^2 d\Omega_4 +dz^2\]
\ee
where $\tilde H_3(r)= 1+\tilde\mu_3/r^3$.
One can easily check that this solution has positive specific heat and
should be stable to small perturbations.
It turns out \cite{Horowitz:1996nw} that the array has larger horizon area than
the homogeneous solution whenever $L$ is larger than $r_0$. (Unlike the
previous cases where we could let $L$ be arbitrarily small, here we require
$L>r_0$.) We are now
in a similar situation as before and can deduce the existence of new
inhomogeneous solutions. The simplest possibility is to connect the
elements of the array with a neutral black 4-brane, i.e. the product of
6D Schwarzschild and $T^4$. The 4-brane wraps the same $T^3$ as the
3-brane, plus the $z$ direction. Since the 
curvature of the near extremal 3-brane is small near the horizon
if the charge is large, 
it should be possible to  construct this
initial data  using the gluing construction. 
By exactly the same arguments as in section 3,
this initial data will evolve, but it cannot settle down to the
homogeneous smeared 3-brane solution since that solution has a 
horizon which is too small. It must settle down to a new smeared black
3-brane, which is near extremal but inhomogeneous in the direction 
perpendicular to the branes. 

Another possibility is to add a ten dimensional
neutral black string to connect the elements of the array. The apparent
horizon would now have the topology of $T^3 \times S^5$ with a $S^7 \times
R$ handle attached. As before, the final topology of the event horizon
could be different, but
it could not settle down to the homogeneous solution. One could also
glue in neutral black 2-branes or 3-branes.

We have started with large near extremal 3-branes. But one could equally
well start with any near extremal brane such that the curvature is small
near the horizon. This includes the M2 and M5 branes in
eleven dimensions. 

\section{Discussion}

We have shown that there exist a large class of inhomogeneous
extended black holes in higher dimension. Unlike the solutions
discussed previously \cite{Horowitz:2001cz}, these solutions exist even when the
translationally invariant black branes are stable. Some of these solutions
remain inhomogeneous even when the length of the compact direction
at infinity goes to zero. As one consequence, this means that
for certain values of the mass and charge, 
there is more than one stable solution. In other words,
the  four dimensional black hole uniqueness theorems  do not extend to 
higher dimensions with horizon topology $S^p\times T^q$.
It should be noted that even for the vacuum Einstein equation, it has
recently been shown that there is no uniqueness theorem. In five
dimensions, in addition
to the rotating black hole, there is another stationary solution
with the same mass and angular momentum which describes
a black string in the shape of a large ring \cite{Emparan:2001wn}.
The ring is rotating
and the centrifugal force balances the gravitational attraction.

The inhomogeneous solutions we have discussed have higher Bekenstein-Hawking
entropy than
their homogeneous counterparts. So one might expect that even though
the homogeneous solution is classically stable, it might quantum mechanically
tunnel to the inhomogeneous state. We cannot yet estimate the probability
for this transition, but expect it to be very small for any macroscopic 
solution.

An important open question is whether one can give a microscopic
description of the entropy for these new solutions.
For many of the homogeneous near-extremal solutions such a description
has been obtained in string theory by taking the limit as the string
coupling $g$ goes to zero \cite{Peet:2000hn}.
The mass and charge remain but no longer gravitate.
So the spacetime becomes flat. The charge is represented by D-branes
and the excess energy is described by fields on the brane. The number
of states turns out to be the exponential of $S_{BH} = A/4$ where
$A$ is the area (in Planck units) of the horizon which appears when
one increases $g$. Is there a similar description for the inhomogeneous
solutions?

The answer is not yet clear. One obvious problem is that we do not yet
have the static inhomogeneous solution explicitly to compute its entropy.
However, given the intuition about what the final state looks like, we
can begin to construct a microscopic model. 
Consider the case of near
extremal 3-branes connected by a neutral thin 4-brane. We expect that
it will settle down to a near extremal 3-brane connected by a smeared 3-brane.
At weak coupling, this corresponds to a large number $N$ of 3-branes all
localized at the same point on the circle, together with a smaller
number $N'$ of 3-branes uniformly spread around the circle.
The states of the localized 3-branes are described, as usual, by a $3+1$
dimensional $U(N)$
gauge theory.
The states of the smeared 3-brane are most easily counted in weak coupling
by applying T-duality \cite{Horowitz:1996nw}. 
Then it becomes a near extremal 4-brane with nontrivial flat connection, and
the low energy excitations are described by a $U(N')$ gauge theory
in $4+1$ dimensions. One might hope that the entropy could be reproduced
by combining both systems. Of course, at this level, one does not see any
sign of inhomogeneity. That must arise from interactions between
these states. One can view the existence of these new supergravity solutions
as giving a prediction for
what happens at strong coupling.

\section*{Acknowledgement}
It is a pleasure to thank J. Isenberg, J. Polchinski,
and S. Ross for useful discussions.
G. H. is supported in part by NSF grant PHY-0070895. K. M. is
supported by a JSPS Postdoctoral Fellowship for Research Abroad.

\end{document}